\documentclass[12pt]{JHEP3}
\usepackage{graphicx}
\newcommand{\be}{\begin{eqnarray}}
\newcommand{\ee}{\end{eqnarray}}
\newcommand{\cl}{\;_2C_l(z)}
\newcommand{\dl}{d_L(z)}

\title{Constraining the expansion history of the universe from the red shift evolution of cosmic shear}
\author{Daniel Levy, Ram Brustein\\
Department of Physics, Ben-Gurion University,
    Beer-Sheva 84105, Israel \\ E-mail: levyda@bgu.ac.il,ramyb@bgu.ac.il}
\preprint{}
\abstract{
We present a quantitative analysis of the constraints on the total equation of state parameter that can be obtained from measuring the red shift evolution of the cosmic shear. We compare the constraints that can be obtained from measurements of the spin two angular multipole moments of the cosmic shear to those resulting from the two dimensional and three dimensional power spectra of the cosmic shear. We find that if the multipole moments of the cosmic shear are measured accurately enough for a few red shifts the constraints on the dark energy equation of state parameter improve significantly compared to those that can be obtained from other measurements.   }
\begin{document}

\section{Introduction}

The observational evidence that the expansion of the universe is being accelerated
by dark energy is strengthening continuously \cite{Kowalski:2008ez,Komatsu:2008hk,Reid:2009xm,Percival:2007yw}.
On the other hand, the theoretical efforts to understand the nature of dark energy and the
underlying reason for the accelerated expansion were less successful so far. It has
become clear that in order to make further theoretical progress better methods of
determining the expansion history of the universe are needed. The kinematic
distance probes in the homogeneous and isotropic universe, such as luminosity
distance ($d_L$), angular distance etc., are limited because they rely on light
emitted from distant sources, and hence measure an integral over the expansion
history \cite{maor1}. Consequently even very precise future measurements of the
kinematic probes will not be sufficient to extract precise model-independent
information on the expansion history. In addition to the kinematic probes one can
use dynamical (fluctuations) probes to help determine the expansion history of the
universe. Cosmological perturbations provide, through their dependence on the
homogeneous and isotropic background,  an independent source of information
\cite{Cooray:2003hd,Linder:2005in,Bertschinger:2006aw}. A promising probe is
the weak gravitational lensing induced by the perturbations. The three
dimensional cosmic shear is a redshift dependent tensor that measures the
shape distortion of distant galaxies as the light emitted from them propagates
through the perturbed universe \cite{Heavens:2003jx,Castro:2005bg}. The two dimensional cosmic
shear has been measured in weak lensing experiments
\cite{wlobs1,wlobs2,wlobs3} where the best cosmological constraints are obtained from \cite{Benjamin:2007ys,Fu:2007qq}. Perliminary studies of 3D analysis of current data have been done \cite{Kitching:2006mq,Massey:2007gh} and planned programs (LSST\cite{Zhan:2008jh}, SNAP/JDEM, DUNE/EUCLID\cite{Refregier:2008js,Refregier:2008nq}) are expected to have some three dimensional capabilities. Several reviews of the theory and observations of weak lensing provide an excellent description of the field
\cite{wlrev2,wlrev4,Munshi:2006fn}.

Previously several investigations on the possible use of the three dimensional cosmic shear to determine the
expansion history of the universe or the nature of dark energy were carried out
\cite{Huterer:2001yu,Hu:2002rm,Jain:2003tb,Bernstein:2003es,Zhang:2003ii,Simpson:2004rz,Upadhye:2004hh,Schimd:2006pa,Heavens:2006uk}.
The standard approach to the analysis of properties of dark energy using weak lensing
relates the shear to the matter power spectrum. Choosing a parametrization for the
dark energy equation of state one can then make numerical estimations for the cosmological parameters,
using the results of simulations to take into account the contribution of the
non-linear regime. Issues concerning the choice of the equation of state parameterizations in this context have
been discussed in \cite{Bassett:2004wz,Douspis:2006rs}. In order to analyze the three dimensional cosmic shear
functional dependence on the dark energy equation of state in a model independent way one needs
observables that can determine in a reliable way the time dependence of the
perturbations' amplitude and not only their spectrum at a fixed redshift. Such
observables can have a simple functional dependence on the expansion history and
hence can provide much more information on its time-dependence than the measurement
of a single spectrum. If the cosmological perturbations are small, the cosmic shear
 depends linearly on them. In this linear regime both observations and
theoretical calculations can be done accurately.

In this paper we analyze the possibility to constrain the total equation of state of the universe
$w_{\rm tot}(z)$ (the ratio of the total pressure of the universe to its energy
density) in a model-independent way by cosmic shear measurements. By ``model-independent" we mean the functional dependence of the cosmic shear on the expansion history of the universe rather than a specific parametrization. In the context of this paper this is equivalent to determining the  sensitivity of the angular spectrum of the three dimensional cosmic shear to changes in the total equation of state parameter.
The question that we wish to pose and answer is: What is the quantitative precision on model-independent information about the expansion history of the universe that can be obtained from cosmic shear measurements with a given accuracy? In particular, we wish to determine whether the angular power spectrum of the three dimensional cosmic shear is more sensitive to changes in the total equation of state parameter than other statistics, and if it is, then by how much. Our results allow us to estimate the accuracy goal needed for shear measurements so they can improve on other accurate tests such as luminosity distance measurements or CMB measurements. Our approach is mostly relevant when trying to estimate the prospects of future lensing surveys for constraining the evolution of the universe in a model independent way.
Our results suggest that precise measurements of the three dimensional cosmic shear at
different redshifts will significantly enhance our ability to constrain the
expansion history of the universe in a model-independent way.

The paper is organized as follows. In Sect.~\ref{theory} we review the theoretical material from~\cite{Levy:2006gs} on which the numerical results, Sect.~\ref{estimate}, will be based. In Sect.~\ref{compare} we compare the cosmological constraints that can be obtained from the angular multipole moments of the three dimensional cosmic shear to those that can be obtained from the 2D and 3D spectra of the cosmic shear and Sect.~\ref{conclusions} contains our conclusions.

\section{Theoretical background}
\label{theory}

In this section we review the theoretical results from~\cite{Levy:2006gs} (where many additional details can be found) on which the subsequent numerical analysis is based.
The three dimensional cosmic shear, $\gamma$, is related to the metric perturbations, $\Phi$, through the integral expression
\be\label{gamma}
  \gamma=\int\limits_0^{w} \!\!dw' \frac{f_K(w-w')}{f_K(w)f_K(w')}
                        \eth \eth \Phi.
\ee
Where $w$ is the radial distance coordinate and $f_K(w)=K^{-1/2}\sin (K^{1/2}w)$ , $w$ or $(-K)^{-1/2}\sinh ((-K)^{1/2}w)$ for a closed $(K>0)$, flat $(K=0)$ or open $(K<0)$ universe respectably, and $\eth$ is the spin-weight operator . See \cite{Levy:2006gs} Section 3 for details.

The metric perturbation is related to the matter density perturbation $\delta_m$ by the Poisson equation
\be
	\nabla^2\Phi=\frac {3 {H_0}^2\Omega_{m_0}}{2} (1+z) \delta_m .
\ee
In the linear regime the solutions for $\delta_m$ are described by a growth function $g(z)$ and the spectrum at some initial given redshift of $\delta_m(z_0)$. In many analyses, one is interested also in the nonlinear regime of $\delta_m$. In this case, one has to use simulations or non-linear approximations and interpolate between the linear and non-linear solutions. We wish to avoid using the non-linear regime as much as possible so we can take advantage of the better accuracy of the theoretical calculations in the linear regime. Previously, some proposals on how to sidestep the problems of modeling the non-linear regime were put forward. It was argued in
\cite{Bernstein:2003es} that one should eliminate the dependence on the growth
factor and concentrate on geometric quantities. A similar approach was followed in
\cite{Zhang:2003ii}. In this case, of course, the measurements will have similar
limitations as the measurements of other kinematic probes (as an example one can compare the resulting constraints in \cite{Taylor:2006aw} with those in \cite{Heavens:2006uk}).

Our approach is different. Since it is the perturbations of the metric that shear really depends on, we solve directly for $\Phi$ rather than for $\delta_m$. While the solution for the growth function with an arbitrary time dependent equation of state can be calculated numerically, the solutions for $\Phi$ have a simple functional dependence on $w_{tot}(z)$. Further,  the linear approximation in this case requires that $|\Phi|\ll 1$ but not necessarily that $\delta\rho_m$ be small. At the epoch of matter-radiation equality the density and metric perturbations are very small, with an amplitude of the order of $10^{-5}$. Through matter domination the metric perturbations are constant while the matter density perturbations grow as the scale factor. Today, $\delta\rho_m$ can reach and on some scales exceed unity.  The metric perturbation $\Phi$, on the other hand,  stays frozen at a small value. The value of $\Phi$ is small also for very small scales, for example,  in a galaxy with $\delta\rho_m\sim~100$ the value of $\Phi$ is less than $10^{-3}$.

For late times, we have to take into account the effects of dark energy. $\Phi$ is no longer constant, however, the solution for $\Phi$ is separable \cite{Levy:2006gs}
\be\label{phi}
	\Phi_+(\eta,\vec{x})=C(\vec{x})\Phi_T(\eta) \;\;,\;\;
				\Phi_T(\eta)= \frac{\sqrt{\rho}}{a} \int
d\widetilde\eta a^2(\widetilde\eta)(1+w_{\rm tot}(\widetilde\eta)).
\ee
where $\eta$ is the conformal time. In red-shift space the time-dependence of the perturbation is thus $\Phi_T(z)$ which has a simple functional dependence on $w_{tot}$ through the first order differential equation
\be\label{phiT}
	 \Phi_T(z)-\frac{2(1+z)}{5+3 w_{\rm tot}(z)} \partial_z\Phi_T(z)& =
\frac{1+w_{\rm tot}(z)}{5+3 w_{\rm tot}(z)}.
\ee

This solution can be used to calculate the growth function for an arbitrary $w_{\rm tot}(z)$.
The solution can also be used to construct from eq.(\ref{gamma}) the spin two angular power spectrum of the shear
\be\label{2cl}
	\;_2C_l(w_1,w_2) = (4\pi)^2 \frac{(l+2)!}{(l-2)!} \int\limits_0^{w_1}\!\!du_1
&& \int\limits_0^{w_2}\!\!du_2 \frac{w_1-u_1}{w_1\;u_1} \frac{w_2-u_2}{w_2\;u_2}
\Phi_T(u_1)\Phi_T(u_2)\!  \cr && \times \int\!\! dq\: q^2 f(q) j_{l}(qu_1)j_{l}(qu_2).
\ee
where $w_{1,2}$ are the distances of the sources and $q$ is the wave number of the perturbation. For a flat spectrum
\be\label{fq}
	f(q)=A\;\left(\frac{2\pi}{q}\right)^3 {T_q}^2(\eta_{in}) .
\ee
${T_q}^2(\eta_{in})$ being the transfer function at early stages of the matter domination epoch (denoted by an initial time $\eta_{in}$) and $A$ is the primordial amplitude. In \cite{Levy:2006gs} we have found that the $\:_2C_l$'s are not sensitive to changes in the shape of the initial spectrum. The evolution with red shift of the $\;_2C_l$ can be evaluated only if we know the distance-redshift relation (written here for a spatially flat universe)
\begin{equation}
\label{lum1}
w=\frac{\dl}{1+z} = \frac{1}{H_0} \int\limits_1^{1+z}\hspace{-.05in}
 d x' \, {\rm exp}
 \left[ - \frac{3}{2} \int\limits_1^{x'}\hspace{-.05in} d \, {\rm
 ln}\, x \, \left(1+ w_{\rm tot}(x)\right) \right].
 \end{equation}
In general, $\dl$, depends on the total equation of state $w_{\rm tot}$,  the spatial curvature $K$ and the value of $H_0$. If the luminosity distance is known to a good accuracy (equal to, or better than, that of the shear measurement) then for a fixed angular scale $l$, $\cl$ is sensitive to changes only in $\Omega_m$ and $w_{tot}(z)$ making it a good candidate for constraining the expansion history of the universe. In what follows we will assume that $\dl$ is known to about a percent level from supernovae observations.

\section{Estimating the cosmic shear angular multipole moments}
\label{estimate}

To see if the $\cl$ can constrain the expansion history of the universe, several questions have to be answered quantitatively. First, we would like to know whether by measuring the $\cl$ it is possible to improve on the kinematical probes and by how much.  Then, we would like to compare the $\cl$ to other statistics that are frequently used in weak lensing analysis. Additionally, since future supernovae observations expect an accuracy of 1\% in determining the luminosity distance, we will see how the constraints on $\Omega_m$ and $w_{tot}$ from knowing $\dl$ to one percent  accuracy compare to those from the red shift evolution of the shear spin two angular power spectrum.

In order to estimate the cosmological constraints we make the following assumptions: That future surveys can measure the $\cl$ at accuracies that are cosmic variance limited for five different red shift bins in the range $0.2<z<1.5$ for about 100 multipoles centered around $l=100$ in the linear regime. Under these conditions we expect that the $\cl$ can be measured at the percent level.

Since each of the tensor $\cl$'s is estimated by $2(l+1)$ independent moments as for the case of regular scalar angular moments, the error estimate follows in a similar way to the error estimate of any Gaussian field measured over a fraction of the sky, $f_{sky}$, with a finite number of galaxies (see, for example, \cite{kaiser} and the appendix of \cite{Knox:1995dq}). Hence, the fractional error for the $\cl$ at a given red-shift is given by
\be\label{clerror}
	\frac{\Delta\!\cl}{\cl}= \sqrt{\frac{2}{(2l+1)f_{sky}}} \left(1+\frac{\sigma_{\gamma}^2}{n_{\mathrm{gal}}(z)\cl}\right)
\ee
 The important difference here is that eq.(\ref{clerror}) is a function of red-shift. The shot noise term includes the intrinsic variance of the shear of a single galaxy $\sigma_{\gamma}^2\sim 0.1$ and the source galaxy density per steradian, $n_{\mathrm{gal}}(z)$, as a function of red-shift. We assume that the source galaxy density distribution is uniform in angle so the total number of galaxies is given by $N_{\mathrm{gal}}= 4\pi n_{\mathrm{gal}}(z)$.

To estimate the error we will look at the lowest red-shift range where the $\cl$ are smallest and the source galaxy density is the smallest so the constraints on the accuracy will be the strongest. For example, for $z\sim 0.2$ $\cl\sim 0.02 \ 2\pi/ l^2$ while for $z\sim 1.5$ $\cl\sim \ 2\pi/ l^2$ \cite{Levy:2006gs}. A standard expected source galaxy density has the form $z^2 e^{-(z/z_0)^2}$ whose median red-shift is somewhat larger than $z_0$. To get the shot noise down to a comparable level to the cosmic variance error for $z=0.2$ one therefore needs $N_{\mathrm{gal}}(0.2)\sim 10\ l^2$ for each angular multipole.
For example, the cosmic variance error of a survey covering half of the sky at $l=100$ is
$\sim 14\%$ and one needs about $10^5$ galaxies to get the shot noise down to this level. To improve on the accuracy of cosmic variance limited measurement one can use binning of multipoles so that the number of independent measurements $\sim (l_{max}^2-l_{min}^2)$. To reach an accuracy on the order of 1\%, about 100 multipoles around $l \sim 100$ must therefore be measured. To reduce the shot noise to the percent level one needs in this case $N(0.2)\sim 10 \left(l_{max}^3/3-l_{min}^3/3\right)\sim 10^7$ galaxies. If the distribution of source galaxies is as given above with median redshift of about $1.25$, one would need about $2\times 10^8$ galaxies. This would be equivalent to a total galaxy source density of about 2 galaxies per squared arcminute.

Whether or not future cosmological observations of the three dimensional cosmic shear and of luminosity distance will achieve the percent accuracy goals that we have described will be determined by future technological advancements in the different fields of observational cosmology. However, at the moment, there does not seem to be any fundamental theoretical or technological reason that indicates that such accuracy goals can not be achieved.

In view of the above, we let the accuracy with which the $\cl$ are assumed to be measured to vary between 1\% and 10\% in our calculations. In addition, as previously mentioned, we assume a prior of 1\% on the redshift-distance relation (i.e. luminosity distance). Since we wish to expose the difference between the sensitivities of $\cl$ and $\dl$, we put strong priors ($1\%$ error) on $K$ and $H_0$ when calculating the sensitivity of $\dl$ which enhances the sensitivity of $\dl$. If we select weak priors the advantage of $\cl$ becomes more pronaunced.

\FIGURE[t]{
		\includegraphics[width=16cm]{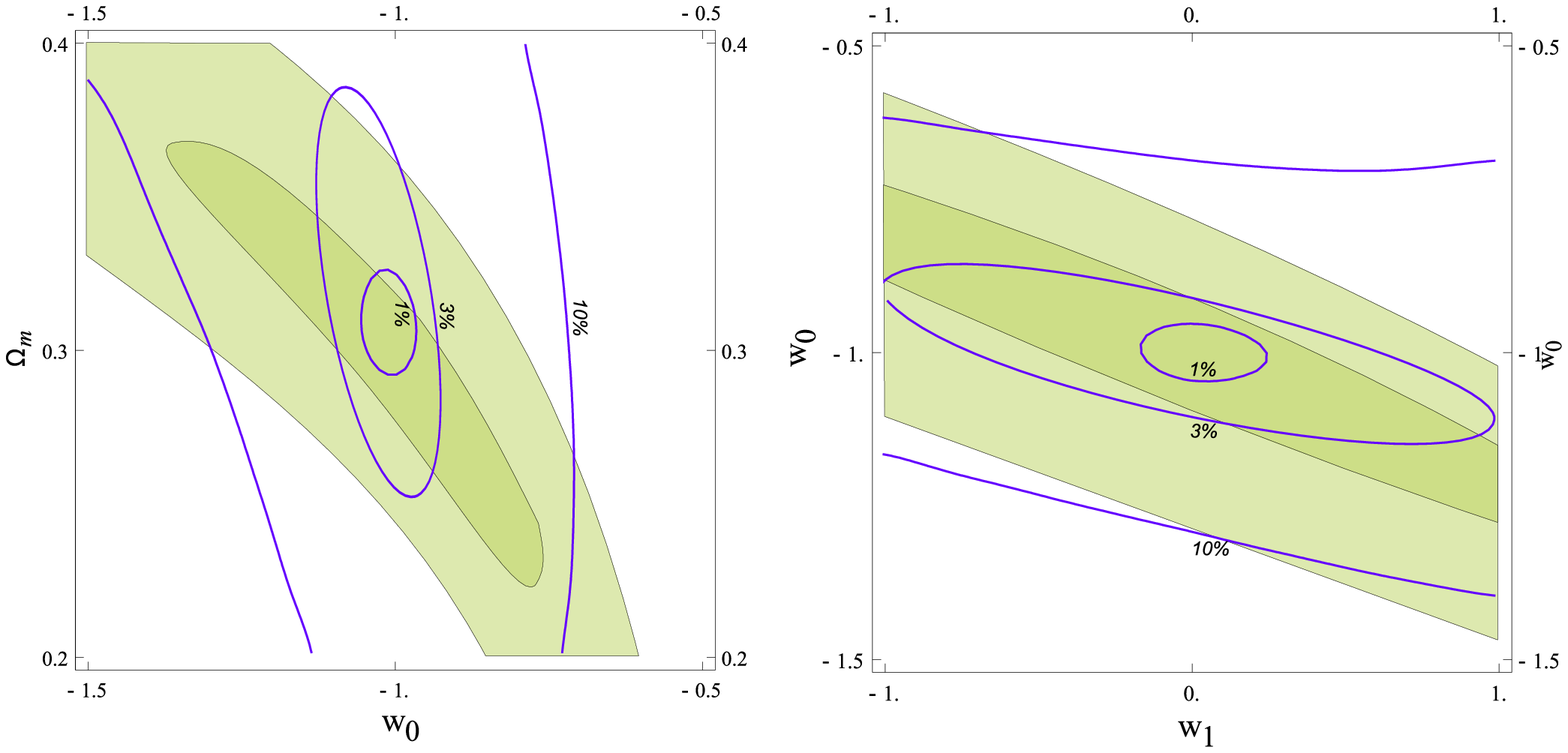}
	\caption{\label{image03}The 1$\sigma$ contours for $\cl$ at $l=100$ with assumed errors of 1\%, 3\% and 10\% (purple lines). The wider filled contours (green) show the 1$\sigma$ and 2$\sigma$ likelihood contours for supernovae $\dl$ measurements with 1\% accuracy. }}

The expression for the $\cl$ as a functional of the total equation of state parameter given in Eqs.~(\ref{phiT}), (\ref{2cl}), does not  assume a particular form of the time dependence of $w_{tot}$ or $w_{DE}$. However, for the purpose of the numerical comparison between the different measures of the cosmic shear we choose, for the moment, a specific parametrization.  We use Linder's parameterization $w_{DE}=w_0+w_1(z/1+z) $, because it is commonly used in the literature and hence its use facilitates the comparison of our results to the results of other theoretical investigations and to most of the existing observational results. That said, we wish to emphasize that our results can be as easily adapted to any other parametrization. In fact, quite a few papers suggest other (improved) parameterizations of $w_{DE}$ that are just as easy to analyze for the measurements of $\cl$. We use the standard likelihood analysis to estimate the constraints on $w_0,w_1$ and $\Omega_m$. For our fiducial model we choose $\Lambda$CDM with $\Omega_m=0.3$ The results can be seen in Figure (\ref{image03}). As we can see, for the $\cl$ to be competitive the measurement accuracy should be 3\% or better. For errors of  10\% the constraints from the shear are weaker than those of the luminosity distance, on the other hand when the error is 1\% the constraints improve significantly.

\FIGURE[t]{
		\includegraphics[width=16cm]{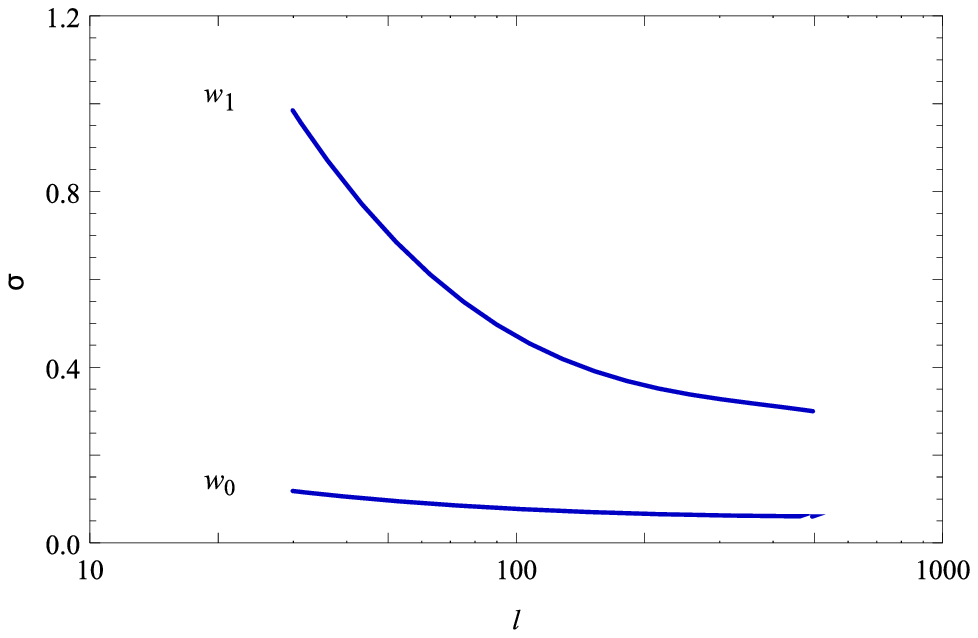}
	\caption{\label{lfunc}The magnitude of the error (1$\sigma$) in $w_0$ (bottom line) and $w_1$ (top) as a function of $l$.} }

A second important aspect of the behavior of the $\cl$ is the improvement of the constraints for higher multipoles. In Fig.~\ref{lfunc} we can see the error in the equation of state parameters decreases as a function of $l$. The reason for that is the fact that the integrand of (\ref{2cl}) samples $\Phi$ at lower redshifts for higher orders of the Bessel function and since $\Phi$ is more sensitive to changes in $w_{tot}$ at the lower redshifts there is more sensitivity at higher $l$'s. This is particularly significant for the $w_1$ parameter which determines the time variation of the equation of state in Linder's parametrization. This behavior is independent of the fact that for higher multipoles the error is smaller and demands only that we use the linear regime.

The range of multipoles that are linear is larger than can be used in the case of the standard two dimensional weak lensing observations. The reason is as follows.
The matter power spectrum is accepted as linear up to $k=0.1 Mpc^{-1}$ which for the two dimensional shear power spectrum is about $l=200$, above which the standard practice is to use non-linear models to estimate it. The metric perturbation $\Phi$, on the other hand, is linear at smaller scales (larger $l$'s). For wavelengths well inside the horizon $\delta\rho_m/\rho_m$ is larger than $\Phi$ by a factor of the order of the square of the ratio of the size of the horizon to the wavelength, $\delta\rho_m/\rho_m\sim(q/H)^2\Phi$. The larger linear range allows us to use the higher multipoles effectively. This is an advantage since as can be seen from Fig.~(\ref{lfunc}), the higher multipoles  are more sensitive to changes in the equation of state and going beyond $l=200$ improves our constraints. The transition to non-linear regime for the matter density is on scales of about  8 Mpc, slightly larger than the scale of clusters. The value for $\Phi$ on the other hand is much smaller even down to scales of 200Kpc, the size of a galaxy. When calculating the $\cl$ the smallest scale that considered here is for the case  $l=500$  (the largest $l$ considered in the paper) at  red shift of $z=0.1$ (the smallest redshift considered in the paper). In this case the relevant scale is about 1.2 Mpc which is still extremely linear in $\Phi$.  As an example, the expected corrections for $\Phi$ from non-linear terms at $k=0.8 Mpc^{-1}$ are of the order of $10^{-7}$.

\section{Shear power spectrum comparisons}
\label{compare}

There are many ways to quantify and measure shear, each having its own merits and drawbacks. Several two dimensional measures of shear have been measured to date and a substantial amount of work on the three dimensional power spectrum has been published. The cosmic shear depends on the metric perturbations and thus like the matter density perturbations is a good source for information on $\sigma_8\;,\Omega_m\;,w_{tot}$ and other cosmological parameters. As previously explained, in the context of cosmological models and explaining the expansion history of the universe, the important point for us is how well it is possible to constrain the total equation of state as a function of red-shift. From this narrow point of view we ask whether different measures of shear perform differently. We will  show that the $\:_2C_l(z)$ is the most sensitive to $w_{tot}(z)$ and gives the best constraints on it. When planning an experiment one questions the accuracy with which any quantity can be measured but we would like to ask a different question: Given an accuracy, what constraints can we obtain on the relevant parameters? There are many different considerations and technical issues that take part in evaluating future measurements and the expected constraints they will give. The answers vary between different surveys and therefore we take this different approach where we try and keep the comparison similar so that the results reflect the sensitivity of the measures and not the quality of the experiments. To that end we choose an accuracy goal which is (almost) optimal so that the results will reflect the best future prospects.

In this section we present a numerical analysis of the sensitivity of different shear measures to changes in the total equation of state. We set an a priori accuracy level of $1\%$ (which is the goal accuracy for future cosmological observations) and analyze the likelihood for $\Omega_m$, $w_0$ and $w_1$ (as in the previous section). The measures we choose are common to all the different methods and differ between them in their intrinsic functional dependence on $w_{tot}(z)$. The first is the two dimensional angular power spectrum \cite{wlrev4}
\be\label{2Dcl}
	P_\kappa (l)=\frac{9 H_0^4 \Omega_0^2}{4 c^4}\int\limits_0^{w_H} dw \frac{\overline{W}^2(w)}{a^2(w)}P_\delta\left(\frac{l}{f_K(w)},w\right) \\
	\overline{W}(w)\equiv \int\limits_w^{w_H} dw' G(w')\frac{f_K(w'-w)}{f_K(w')}, \nonumber
\ee
where $G(w)$ is the normalized source distance distribution function.
The second is the three dimensional power spectrum \cite{Castro:2005bg}
\be\label{3Dcl}
	C_l^{\gamma\gamma}(k_1,k_2) =\frac{4}{\pi^2 c^4}\frac{(l+2)!}{(l-2)!}
																\int\limits_0^\infty k^2 dk I_l(k_1,k) I_l(k_2,k) \\
	I_l(k_i,k)\equiv k_i \int\limits_0^\infty dr r j_l(k_i r)\int\limits_0^r dr' \left(\frac{r-r'}{r'}\right)
								 j_l(k r')\sqrt{P^{\Phi\Phi}(k;r')}, \nonumber
\ee
and the third is the $\;_2C_l(z)$.
\FIGURE[t]{
		\includegraphics[width=16cm]{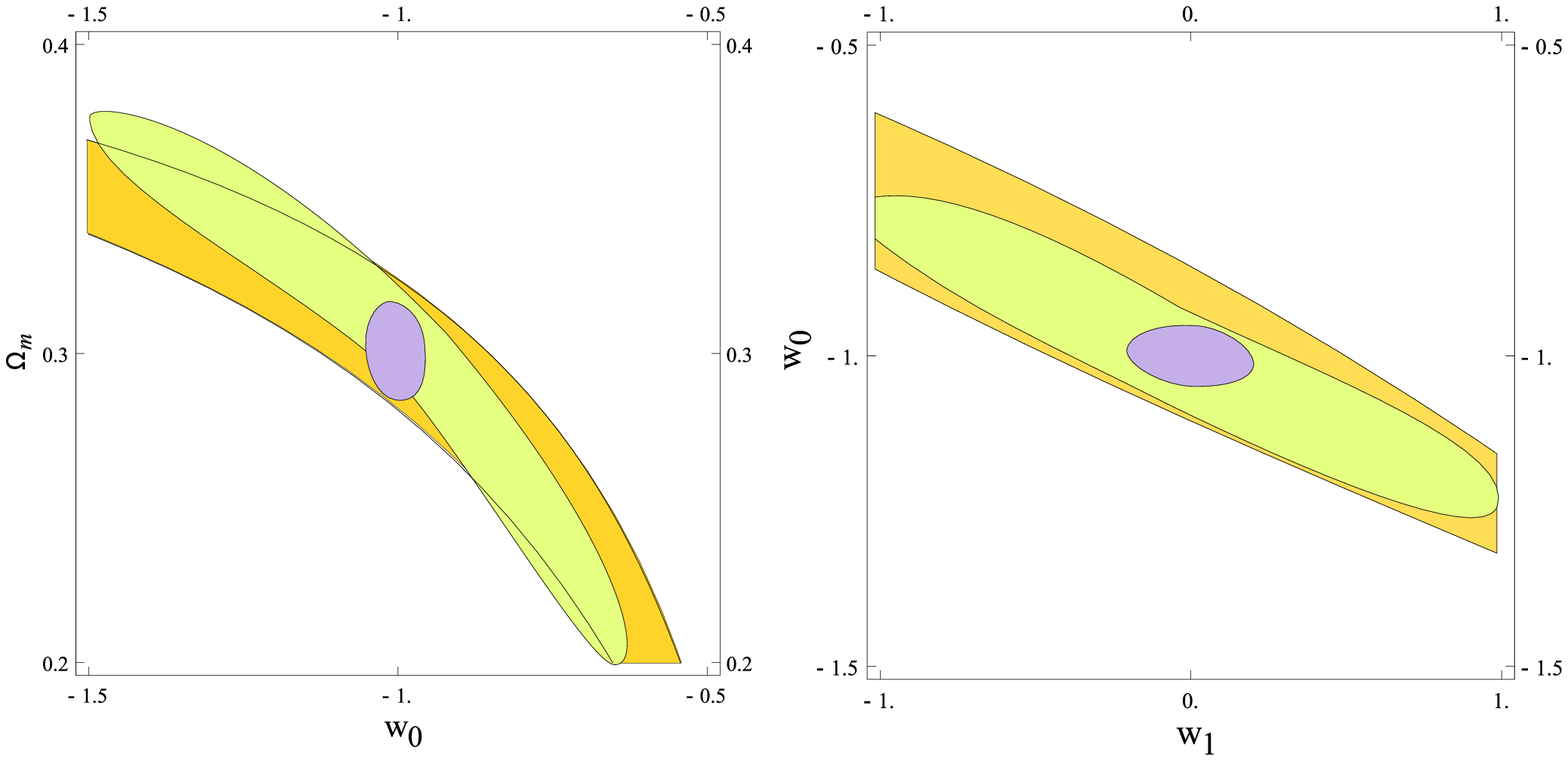}
	\caption{\label{image5}The 1$\sigma$ contours of $\;_2C_l(z)$ (purple),$C_l^{\gamma\gamma}(k_1,k_2)$ (green) and $P_\kappa (l)$ (orange).}}

The assumptions concerning the distance red-shift relation and all the priors are similar to those in the previous section. In addition, these quantities depend on the normalization of the power spectrum. Although it is common to use $\sigma_8$, the metric perturbation is proportional to the primordial amplitude of perturbations, $A$. The value of $A$ is obtained from CMB measurements to $5\%>$ accuracy. The $\cl$ are insenstive to the value of $A$ and assuming a prior of $5\%$ then marginilizing over it makes no differnce to the $w_0,w_1$ likelihood. For the 2D and 3D spectrums on the other hand margenilizing over $A$ increases the contours. Therfore, in the following calculations we keep the value of $A$ fixed, so that it is the difference in sensitivity to $w_{tot}$ that is demonstrated.

We can see from the contours in Fig.\ref{image5} that the $\;_2C_l(z)$ is more sensitive to changes in $w_{tot}$ and $\Omega_m$, the reason for this is clear upon examination of the functional dependence of the different measures on these parameters. The dependence of perturbations on the total equation of state comes through the growth function. The two dimensional power spectrum is a projection, therefore it contains a double integral over $\Phi_T(z)$ and the result is degenerate with respect to $w_{tot}$. This is solved by measuring distance (red shift) dependent quantities such as the three dimensional spectrum. But, although the full spectrum measures best all the cosmological parameters of the model it measures the shape and evolution of the spectrum simultaneously and therefore is less effective in distinguishing between the information from the growth function and from the spectrum's shape. The $\;_2C_l(z)$ on the other hand is insensitive to the $k$ dependence of the spectrum and therefore can not determine $\sigma_8$ directly, but it has a straightforward functional dependence on $w_{tot}$ and $\Omega_m$ and therefore gives better constraints on them.

For the three dimensional methods we have assumed that a single (binned)
multipole is measured but both quantities can be measured at several (binned) multipoles and with different correlations ($C_l^{\gamma\gamma}(k_1,k_2)\;,\;k_1\neq k_2$ or $\;_2C_l(z_1,z_2)\;,\;z_1\neq z_2$) thus increasing the final accuracy. This is important since for a given survey magnitude the error on $P_\kappa (l)$ should be smaller than on $\cl$ because of the higher number density of galaxies, but as the number of redshifts grows the number of cross correlations grows too and with it the accuracy.

\subsection*{Multi parameter modeling of $w_{DE}$}

The main advantage of using the $\cl$ is their explicit and fully functional dependence on $w_{tot}$ without prior assumptions on the functional form of $w_{\rm tot}$. The major improvement of using them compared to other approaches is in their ability to place constraints on the evolution of $w_{tot}(z)$. Since fundamental theory offers little insight  as to the functional form of the dark energy equation of state, and to demonstrate the sensitivity of the $\cl$ to changes in it,  we choose a multi dimensional parameterization to further explore the difference between the various measures that we have mentioned above. We choose a parametrization of $w_{DE}$ \cite{Albrecht:2007qy,Albrecht:2009ct}  in which the dark energy equation of state is parametrized by a piecewise constant function. The constants $w_i$ are the constant values of $w(a)$  in bins of $\Delta a$. We choose (for practical reasons) to divide the range $1 < a < 0.2$ into 5 bins of width $\Delta a=0.16$, so that in redshift space $z_i+1=1/a_i$ we have
\be
	w_{DE}(z)=\sum_{i=0}^{4} w_i T(z_i,z_{i+1}).
\ee
The top-hat function $T(z_i,z_{i+1})$ is equal to unity for $z_i>z\geq z_{i+1}$ and vanishes otherwise.
In Fig.(\ref{fig5bin}) we present the $1\sigma$ errors for each of the parameters marginalized over $\Omega_m$. The calculations where done under the same assumptions as in the previous sections. This is a somewhat simpler analysis than that of the "principal components"\cite{Albrecht:2007qy} but it is sufficient to demonstrate the relative difference between the three measures of cosmic shear. We can see once again that in the first 4 bins the $\cl$ is significantly more sensitive to changes in $w_{DE}(z)$ than the other methods mentioned above. We can also see once again that the luminosity distance, even when determined to one percent, cannot sufficiently constrain the equation of state. In the fifth bin, $w_4$ gives the value for $z>1.7$ which is beyond the assumed measured range (5 measurements in the range $0.2<z<1.5$) and therefore the constraints are very weak. Comparing our results to existing multi-parameter analyses \cite{Albrecht:2007qy,Huterer:2000mj,Huterer:2002hy,Knox:2004vw,bridle} the most pronounced difference is the low red-shift sensitivity. A possible origin of this difference is that the expression of the angular moments contain the square of the galaxy distribution function (see, for example, \cite{Simpson:2004rz} eqs.~(2),(3)) which vanishes for low redshifts. This issue should be better understood and we leave it for a future investigation.  For higher redshifts, the value of $w$ is constrained by the $\cl$ better than the other statistics and the relative advantage of the $\cl$ increases with redshift.
One may wonder why is it that the $\cl$ do better than $C_l^{\gamma\gamma}(k_1,k_2)$ since they contain essentially the same information. The reason is that the redshift evolution of the $\cl$ measures the growth function directly while which $C_l^{\gamma\gamma}(k_1,k_2)$ does not. Measuring the growth function is what gives us the better constraints on the equation of state of the dark energy specifically.  Similar effects can be obtained by cross correlating $P_\kappa (l)$ at different redshift bins.

\FIGURE[t]{
		\includegraphics[width=16cm]{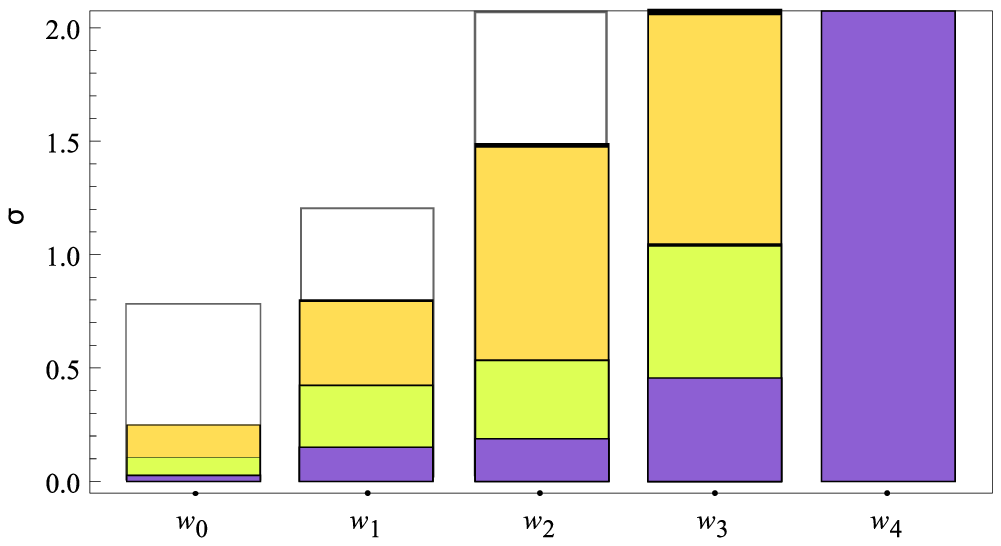}
	\caption{\label{fig5bin}The 1$\sigma$ magnitude of the five redshift bin values of $w_{DE}(z)$ marginalized over $\Omega_m$ for $\;_2C_l(z)$ (purple),$C_l^{\gamma\gamma}(k_1,k_2)$ (green) and $P_\kappa (l)$ (orange). The empty bars represent the 1$\sigma$ magnitude for $\dl$ assuming it is measured at 1\% accuracy. The redshift bins are delimited by $\{0,\ .19,\ .47,\ .92,\ 1.78,\ 4.0$\}.}}

We can observe  that the $\cl$'s are indeed more sensitive to a deviation of  $w_{DE}$ from $-1$ as in the case of a cosmological constant and that the extra sensitivity is more significant at the higher redshifts. Thus, the three dimensional shear measurements, and specifically the red shift evolution of the spin two angular power spectrum, are an important tool in determining the evolution of the equation of state.

\section{Conclusions}
\label{conclusions}

In this paper (as well as its predecessor \cite{Levy:2006gs}) we have clarified the dependence of the cosmic shear on the expansion rate of the universe and have made a preliminary quantitative analysis of its sensitivity to changes in the dark energy equation of state. Theoretical arguments suggest that the red shift evolution of shear spin weight two angular power spectrum multipole moments $\cl$'s,  are the best statistics for this purpose. They have the simplest dependence on the metric perturbations $\Phi$, and provide a more direct connection between the measured quantities and the theoretically interesting quantities. More specifically, theoretical arguments suggest that precise knowledge of the $\cl$'s should allow to obtain better constraints on $w_{tot}(z)$ and the expansion history of the universe. We have demonstrated the validity of this suggestion by a preliminary numerical analysis.
To evaluate the magnitude of improvement that we should expect from $\cl$ measurements on future surveys a full scale numerical analysis of currently planned programs should be carried out.

Comparing the constraints that can be obtained by using the $\cl$ and other measures of cosmic shear, such as the 2D and 3D power spectrum as well as kinematic distance probes such as $\dl$, we see that even when all other parameters are kept fixed the $\cl$'s are more sensitive to changes in $w_{DE}$ and therefore better for putting constraints on $w_{\rm tot}(z)$. Furthermore, we have shown that up to the maximal red-shift at which $\cl$ is measured, it is possible  to get a significant improvement on the constraints of $w_{\rm tot}$. This fact is important when one is trying to determine whether the dark energy is  a cosmological constant or some other dynamical source. The constraints on the current value of $w_{DE}(0)=-1$ are fairly tight so the expected improvement from future observations in the Figure of Merit, for example, will come mostly from better constraints on the time evolution of the dark energy equation of state parameter. Therefore, the best prospect for deciding whether the dark energy is a cosmological constant or dynamical lies in better constraints on $w_{tot}(z)$ at higher redshifts, such as those that can be obtained from measuring the $\cl$'s.

\section*{Acknowledgments}
We thank Marcello Cacciato and Ofer Lahav for interesting discussions and useful comments.
This research is supported by The Israel Science Foundation grant no 470/06.

%%%%%%%%%%%%%%%%%%%%%%%%%%%%%%%%%%%%%%%%%%%%%%%%%%%%%%%%%%%%%%%%%%%%%%%%%%%%%%%%%%%%%%%%%%%%%%%%%%%%%%%%%%%%%%%%%%%%

\end{document}